\documentclass[aps,10pt,prb,twocolumn,showpacs,superscriptaddress]{revtex4-1}

\usepackage[colorlinks]{hyperref}	
\usepackage{amsfonts}
\usepackage{epsfig}
\usepackage{bm}
\usepackage{amsmath}
\usepackage{subfigure}
\usepackage{grffile}  

\newcommand{\vc}[1]{\boldsymbol{{#1}}}
\newcommand{\rmd}{\mathrm{d}}	
\newcommand{\C}{\mathcal{C}}	
\newcommand{\change}[1]{#1}

\begin{document}

\title{Electron pairing in periodic potentials under an external electric field}

\author{C. Gaul}

\affiliation{GISC, Departamento de F\'{\i}sica de Materiales, Universidad
Complutense, E-28040 Madrid, Spain}
\affiliation{CEI Campus Moncloa, UCM-UPM, Madrid, Spain}

\author{A. Rodr\'{\i}guez}

\affiliation{CEI Campus Moncloa, UCM-UPM, Madrid, Spain}
\affiliation{GISC, Departamento de Matem\'{a}tica Aplicada y Estad\'{\i}stica, 
Universidad  Polit\'{e}cnica, E-28040 Madrid, Spain}

\author{R. P. A. Lima}

\affiliation{Instituto de F\'{\i}sica, Universidade Federal de Alagoas,
Macei\'{o}, AL 57072-970, Brazil}

\author{F. Dom\'{\i}nguez-Adame}

\affiliation{GISC, Departamento de F\'{\i}sica de Materiales, Universidad
Complutense, E-28040 Madrid, Spain}
\affiliation{CEI Campus Moncloa, UCM-UPM, Madrid, Spain}

\date{June 26, 2013}

\begin{abstract}

We study the semiclassical dynamics of interacting electrons in a biased crystal
lattice. A complex dynamical scenario emerges from the interplay between the
Coulomb and the external electric fields. When the electrons are far apart, the
Coulomb potential may be small compared to the external potential and the
electrons oscillate with effective Bloch frequencies, determined by the local
electric field. In the opposite case, nearby electrons either separate or form a
bound pair, depending on the initial energy compared to the band width. The pair
due to the Coulomb field is stable even in the absence of the external field. 

\end{abstract}


\maketitle

\section{Introduction}

The dynamics of quantum electrons in solids subjected to a uniform electric
field is rather non-intuitive. According to the semiclassical picture 
introduced by Bloch\cite{Bloch1929} and Zener,\cite{Zener1934} noninteracting
electrons do not accelerate uniformly in real space but oscillate instead. These
coherent oscillations are known as Bloch oscillations~(BOs). Much after their
theoretical prediction, electronic BOs were observed in semiconductor
superlattices.\cite{Feldmann1992,Leo1992,Leo1998} BOs persist until electrons
lose their phase coherence through scattering processes. Among the various
scattering processes that may affect the coherent motion of carriers,
electron-electron interactions have their own peculiarities. In this regard,
Freericks has studied the dynamics of conduction electrons and localized
electrons, which do not move but interact with the conduction electrons when they
are in the same unit cell.\cite{Freericks2008} It was shown that BOs are sharply
damped and become quite irregular in time in this case. Interaction between
conduction electrons is expected to have less impact since all electrons
oscillate with the same frequency.  Nevertheless, Hubbard-like interactions
between particles in the same band also induce the irreversible decay of
BOs.\cite{Buchleitner2003}

Several works have explored the problem of few particles in the BO regime.
\change{
The possibility of fractional period in the collective dynamics of several
coupled quasi-particles was predicted in a series of papers devoted to BOs of
magnetic solitons in inhomogeneous magnetic 
fields.\cite{Kosevich1998_PhysicaD,Kosevich1998,Kosevich2001} In particular, it
was shown that if a soliton binds $N$ excitations, its BO frequency is
proportional to $N$. More recently,
}
Khomeriki \emph{et al.}\ studied the dynamics of few interacting bosons in a
periodic lattice and subjected to a constant force.\cite{Khomeriki2010} They
found that for strong interaction the BO regime re-emerges with fractional Bloch
periods, which are inversely proportional to the number of bosons clustered into
a bound state. The dynamics of two interacting electrons was discussed by Claro
\emph{et al.}\ within the framework of the Hubbard Hamiltonian.\cite{Claro2003}
They concluded that electron-electron interaction induces time-dependent
oscillations whose period depends on the strength and range of the coupling
only. The dynamics of the electron pair without long-range interaction also
depends on the initial conditions. When initially they are far apart, the
dynamics is that of the single particle BO, as expected.\cite{Dias2007} On the
contrary, a period doubling is found when the two electrons remain close,
indicating that the pair behaves effectively as a composite
particle.\cite{Dias2007}

Usually, BOs in the correlated regime are studied with contact interaction in
the Hubbard Hamiltonian.\cite{Longhi2011,Dias2010}
\change{The tight-binding single-band description is a good approximation if the
band-gap frequency, i.e, the band gap divided by $\hbar$, is much larger than the
Bloch frequency due to the external field.\cite{Hone1996} The approximation of
short-range interactions} is well justified for describing two electrons
interacting by a screened Coulomb potential when the screening length is smaller
than the lattice spacing. However, its applicability is questionable when the
screening length is large. Although long-range interactions can be implemented
in the Hubbard Hamiltonian,\cite{Claro2003} the resulting equations are
complicated, even for two electrons only. In this paper we consider two
interacting electrons in a lattice subjected to a constant electric field. We
study their dynamics within the semiclassical framework when they interact by
the Coulomb potential. We identify new oscillation regimes, that were missed in
previous studies, due to the long-range nature of the Coulomb potential.  In
particular, we find that two electrons form a bound pair if the energy of the
relative motion exceeds \change{the upper band edge},
even in the absence of external field.
\change{
It is worth mentioning that the possibility of electron pairing in solids caused by
the repulsive Coulomb interaction was already pointed out in a footnote of the
textbook by Lifshitz and Pitaevskii.~\cite{Lifshitz1989} One of our aims is to
elaborate this idea and to present a detailed analysis of the conditions needed
to form the pair.
}

\section{Semiclassical approach} \label{semiclassical}

The semiclassical dynamics of an electron in a periodic lattice is solely
parameterized by its central position $\vc r$ and its central  momentum $\hbar
\vc k$. Thus, the equations of motion for two independent electrons are
$\hbar\dot{\vc k}_{i} = -e\vc{\mathcal E}$ ($i=1,2$), where $\vc{\mathcal E}$ is
the applied electric field. The group velocity of the electrons is given by 
\begin{subequations}\label{equations_of_motion}
\begin{equation}
\dot{\vc r}_i\equiv {\vc v}_\mathrm{g}({\vc k}_i) = \frac{1}{\hbar}\,
\frac{\partial E({\vc k}_i)}{\partial {\vc k}_i}\ , \qquad i=1,2\ .
\label{group_velocity}
\end{equation}
Within the tight-binding approximation, the dispersion relation of the simple
hypercubic lattice is given by $E({\vc k}_i) =
-2J\sum_{\mu=1}^d\cos(k_{i,\mu} a)$, where $a$ is the lattice constant, $d$ is the spatial dimension of the
lattice, and we assume $J>0$ hereafter.

When the Coulomb repulsion between the electrons is taken into account, the
local electric field is the external electric field $\vc{\mathcal E}$ plus the
Coulomb field from the other electron, which results in the equation of motion 
\begin{equation}
\hbar \dot{\vc k}_i=-e\vc{\mathcal E}+\frac{e^2}{\epsilon}\,
\frac{{\vc r}_i-{\vc r}_j}{|{\vc r}_i-{\vc r}_j|^3}\ , \qquad i\neq j\ ,
\label{momenta}
\end{equation}
\end{subequations}
where $\epsilon$ is the dielectric constant of the solid. Since the interaction
term depends only on the relative coordinate, it is appropriate to make a
canonical transformation to total and relative coordinates and quasimomenta.
Thus, we introduce ${\vc r}={\vc r}_1-{\vc r}_2$, ${\vc R}=({\vc r}_1+{\vc
r}_2)/2$, ${\vc k}=(1/2)({\vc k}_1-{\vc k}_2)$  and ${\vc K}={\vc k}_1+{\vc
k}_2$. In order to work in more convenient dimensionless units, we make the
substitutions $t\to J t/\hbar$, $\vc R\to {\vc R}/a$, $\vc r\to {\vc r}/a$, 
$\vc K\to a{\vc K}$, $\vc k\to a\vc k$ to get 
\begin{subequations}
\label{relative_equation_of_motion_dimensionless}
\begin{eqnarray}
\dot{R}_\mu&=&2 
\sin\left({K_\mu}/{2}\right)\cos(k_\mu)\ ,
\label{total_position}\\
\dot{r}_\mu &=&4
\cos\left({K_\mu}/{2}\right)\sin(k_\mu)\ , 
\label{relative_position}\\
\dot{\vc K} &=& -2\,\vc{F} \ ,
\label{total_momentum}\\
\dot{\vc k} &=& g_c\,\frac{\vc r}{r^3} \ ,
\label{relative_momentum}
\end{eqnarray}
\end{subequations}
where the dimensionless magnitudes $\vc{F}\equiv e a\vc{\mathcal E}/J$ and
$g_c\equiv e^2/(J a\epsilon)$ have been introduced. The solution of
Eq.\ \eqref{total_momentum} is trivial and the result can be inserted in
Eqs.\ \eqref{total_position} and~\eqref{relative_position} to reduce the
number of equations.

To estimate the range of validity of the equations of motion
\eqref{relative_equation_of_motion_dimensionless}, one has to take into account
that in reality each electron is represented by a wave packet with a finite
width $\sigma$, which has to be much larger than the lattice constant $a$, but
smaller than the separation of the wave packets $|\vc{r}_1-\vc{r}_2|$. In terms
of the dimensionless coordinates, the validity condition is $|\vc{r}| \gg 1$.

As mentioned before, the condition for the single-band description is
that the band-gap frequency well exceeds the effective Bloch
frequency.\cite{Hone1996} In the dimensionless units of 
Eq.~\eqref{relative_equation_of_motion_dimensionless}, this condition implies
that both $g_{\rm c}/r^2$ and $F$ should be smaller than the band-gap divided by
$J$. Thus, again, the approximation is bound to fail should the particles get
too close to each other or if the external field is too strong.

\section{Zero external field}\label{secZeroField}

Let us consider first the simple case when the external field is absent and
restrict the discussion to one dimension for the time being. 
According to Eq.\ \eqref{total_momentum}, $K=K_0$ is a constant of motion.
Equations~\eqref{relative_position} and~\eqref{relative_momentum} then form a
closed set of equations governed by the Hamiltonian of the relative motion
\begin{equation}
 H_{\rm r} = \frac{g_c}{|x|} -\lambda_0 \cos k\ ,
 \label{eqRelHamiltonian}
\end{equation}
where $\lambda_0 = 4 \cos(K_0/2)$. 
\begin{figure}[tb]
\includegraphics[width=\linewidth]{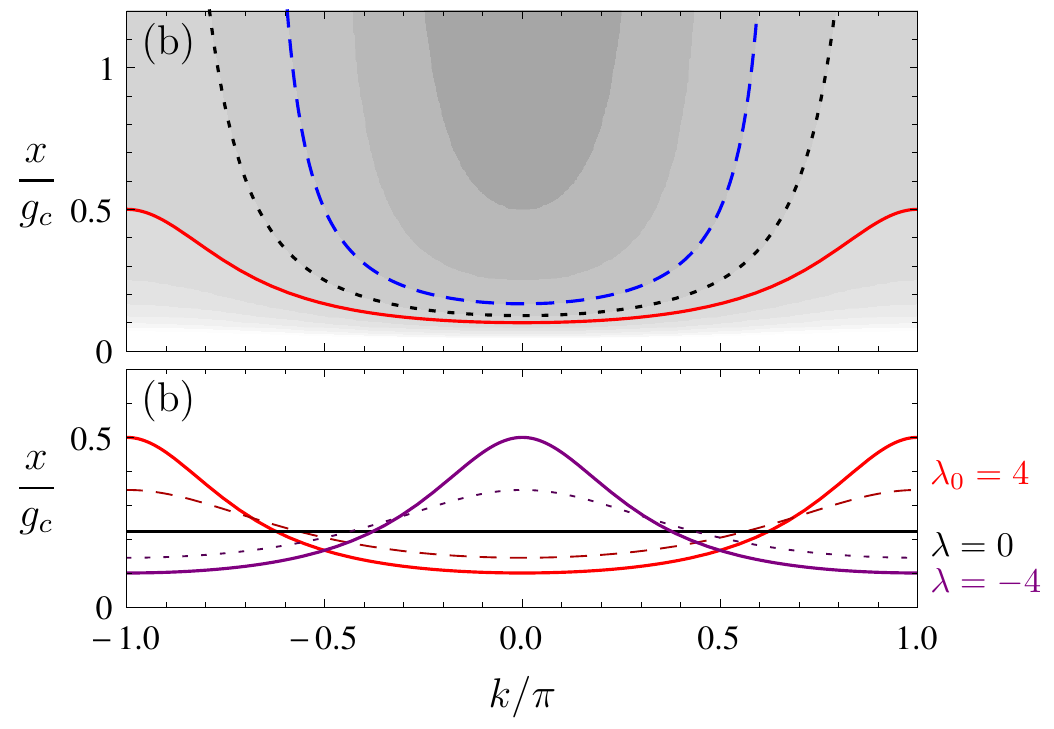}
\caption{(Color online) (a) Phase-space trajectories in absence of external
field for $\lambda_0=4$, given as energy contours of Hamiltonian
\eqref{eqRelHamiltonian} (dark: low energies; light: high energies).
Trajectories from  the unbound regime ($\C = 0.5$, blue dashed line), the
oscillating regime ($\C = 1.5$, solid red line) and the separatrix $\C=1$ (black
dotted line). (b) Phase-space trajectories for adiabatically varying parameter
$\lambda$. The curve for $\lambda_0=4$ is the same as the solid red curve in
panel (a), i.e., $E_0 = 1.5 \lambda_0$. As $\lambda$ changes, the energy adjusts
itself according to Eq.\ \eqref{eqEr}, conserving the phase-space volume under
the curve. For $\lambda=0$ the oscillation comes to a halt at $x/g_c =
\big(E_0^2 - \lambda_0^2\big)^{-1/2}  \approx 0.223$.}
\label{fig1_phasespace}
\end{figure}
The phase-space trajectories of the relative motion are given by the contour
lines $H_{\rm r}=E_0$. These trajectories can be classified according to the
value of the parameter $ \C = E_0/|\lambda_0|$, i.e., the energy of the relative
motion compared to the \emph{effective} \change{upper band edge or} half band
width $|\lambda_0|$. As illustrated in Fig.~\ref{fig1_phasespace}(a), there are
two qualitatively different regimes.  If $\C < 1$, trajectories are
\emph{unbounded\/} in $x$ (blue dashed lines in the plot). On the contrary, for
$\C > 1$ the trajectories are \emph{bounded} in $x$ (red solid line in the
plot), thus corresponding to oscillatory solutions. The curve defined by $\C=1$
is the separatrix (black dotted line in the plot) between the two regimes.
\change{
In this context it should be mentioned that a similar separatrix was already
described and experimentally studied in the search for coherent Hall effect in
semiconductor superlattices subjected to crossed electric and magnetic
fields.\cite{Kosevich2001b,Bauer2002,Hummel2005} In these works it was found
that BOs are suppressed at high magnetic field  and the motion of a single
electron in real space corresponds to a non-oscillatory drift.
}

\begin{figure}[tb]
\includegraphics[width=0.95\columnwidth,clip,angle=0]{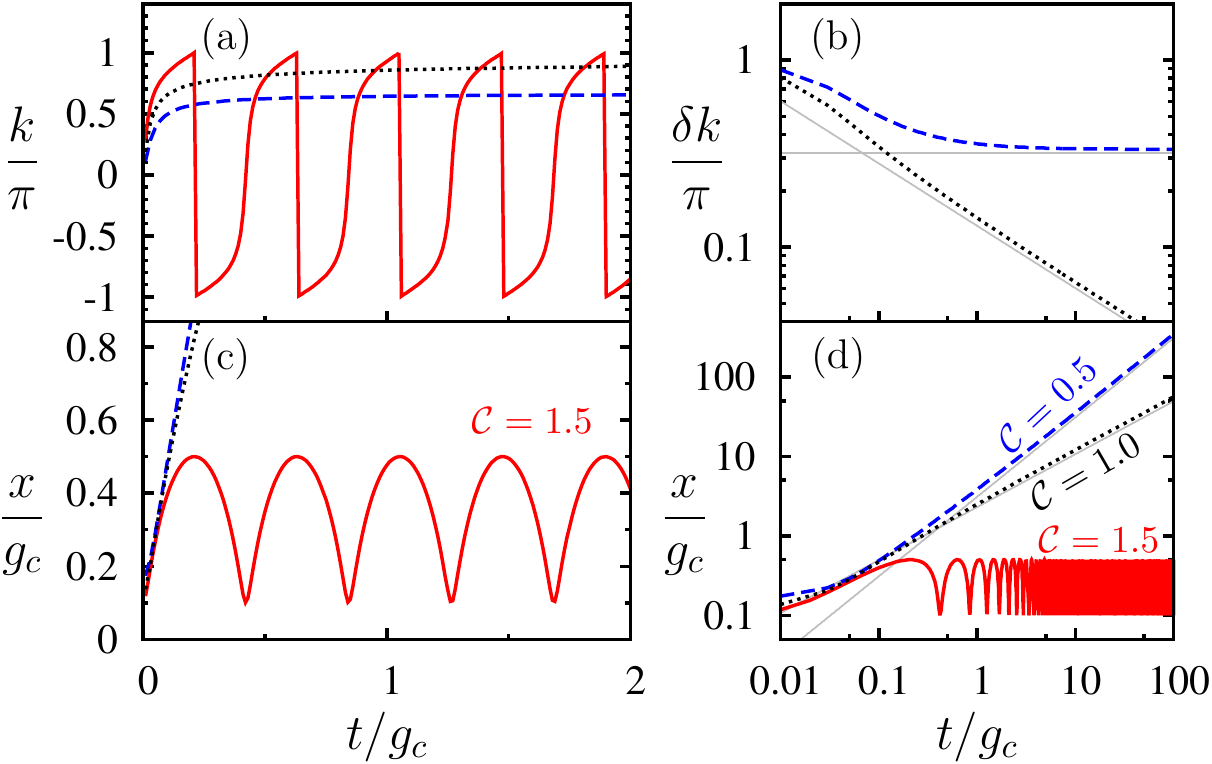}
\caption{(Color online) Time evolution of the relative momentum $k$ [panels (a) and (b)] and the particle separation $x$ [panels (c) and (d)] in the different regimes of Fig.\ \ref{fig1_phasespace}(a) with the same color code.
Panel (b) shows $\delta k = \pi - k(t)$ for the unbound and the separatrix case on a logarithmic scale,
the gray lines indicating the asymptotic behavior $\delta k \sim \text{cst.}$ and $\delta k \propto t^{-1/3}$, respectively.  
Panel (d) shows $x(t)$ and and the asymptotic behaviors $x(t) \sim t$ and $x(t)\sim t^{2/3}$ for the unbound and separatrix cases, respectively.
}
\label{fig_2}
\end{figure}

Figure~\ref{fig_2} shows the time evolution of the three trajectories highlighted in Fig.\ \ref{fig1_phasespace}(a), obtained from the numerical solution of Eqs.~\eqref{relative_momentum} and~\eqref{relative_position} for $K_0=0$, 
$k_0=0$ and three values of the initial separation, namely
$g_c/x_0 = 6, 8, 10$, resulting in $\C = 0.5, 1.0, 1.5$, respectively. 
%
In the unbound regime $\C < 1$, the relative momentum never reaches the edge of the Brillouin zone (BZ), but converges to a value smaller than $\pi$. This results in a finite group velocity and a ballistic separation of the two particles [see blue dashed line in Fig.\ \ref{fig_2}(d)].
This behavior is similar to the dynamics of two electrons in a uniform medium, and no signatures of BOs are found, although the electrons move in a periodic lattice under an electric Coulomb field. 

The oscillations in the case $\C > 1$ are anharmonic [Fig.\ \ref{fig_2}(c)] but they are similar to standard BOs, in the sense that they are driven by an electric field (in this case, by the Coulomb field due to the other electron) and that the relative momentum reaches the edges of the BZ [Fig.\ \ref{fig_2}(a)]. 
Therefore, the periodic potential and the Coulomb \emph{repulsion\/} between the two electrons are responsible for their pairing.
When $\C$ approaches unity from above (bounded trajectories), both the period and the oscillation amplitude tend to infinity;
the \emph{separatrix} [black dotted line in Figs.\ \ref{fig1_phasespace}(a) and \ref{fig_2}] corresponds to 
$\C =1$. Its asymptotic behavior is characterized by $k = \pi - \delta k$, where $\delta k \sim t^{-1/3}$.
Thus, $k$ comes to rest just at the edge of the BZ, where the group velocity vanishes.
The asymptotic behavior of the particle separation is $x(t) \sim t^{2/3}$, i.e., sub-ballistic [see black dotted line in Fig.\ \ref{fig_2}(d)].

Only when the initial conditions satisfy $E_0 > \lambda_0$, i.e., the initial energy is too high to be converted completely into kinetic energy, the dynamics displays oscillations in $x$ (paired electrons) and an unbounded increase of $k$.

After having solved the dynamics of the relative motion, one solves the equation
of motion~\eqref{total_position} for the center of mass. The special case
$K_0=0$ yields $X(t)=X(0)$, so the position and momentum of the center of mass
remain constant, as expected.

\section{Pairing in the adiabatic regime}
We now consider a weak external field, such that $F x_0$ is the smallest of all energy scales, i.e., $F x(t) \ll 1,\,  g_c /x(t)$. 
A perturbative approach, however, is not possible because on long
time scales $K$ grows without bounds and $4 \cos(K/2) = 4 \cos(F t)$ performs
full oscillations. We can, however, consider the adiabatic regime, where
$\cos(Ft)$ varies on a much longer time scale than the dynamics of $x$ and $k$.
In other words, $\cos(Ft)$ can be considered constant during one cycle of $x$ and
$k$. Thus, in the adiabatic limit we can safely replace $4 \cos(Ft)$ by  a
constant $\lambda$. For a given value of $\lambda$, the phase-space trajectories
are given as equipotential lines of $H_\mathrm{r}$. Consequently
\begin{equation}
x(k)=\frac{g_c}{E_{\rm r} + \lambda \cos k} \ ,
\label{trajectory}
\end{equation}
where $E_{\rm r}$ is the energy of the relative motion.

We are interested in the paired regime $E_{\rm r}>\lambda$ [see solid red lines in
Fig.~\ref{fig1_phasespace}(a) and Fig.~\ref{fig_2}]. As the parameter
$\lambda$ in the Hamiltonian changes slowly, the phase-space trajectories are
deformed in time. The energy $E_{\rm r}$ of the relative motion is not conserved
because of the time dependence of $\lambda$. However, the phase-space area
enclosed by a trajectory is an \emph{adiabatic invariant}~\cite{LandauMechanics}
\begin{equation}
  \int_{-\pi}^{\pi} \rmd k \, x(k) 
= \frac{2\pi g_c}{\sqrt{E_{\rm r}^2 - \lambda^2}} 
= \text{constant} \ .
\label{eqAdInv}
\end{equation}
Thus, as $\lambda$ deviates from its initial value $\lambda_0=4$, 
the energy of the relative motion varies as 
\begin{equation}\label{eqEr}
 E_{\rm r}(\lambda) = \sqrt{E_0^2 - \lambda_0^2 + \lambda^2} \ .
\end{equation}
In Fig.~\ref{fig1_phasespace}(b), phase-space trajectories are shown for different values
of $\lambda$ 
with $E_{\rm r}$  according to~\eqref{eqEr}, such that
the phase-space area~\eqref{eqAdInv} is constant. As $\lambda$ decreases from
4 to zero, the amplitude of the (anharmonic) oscillation vanishes.  For
negative values, the oscillation is inverted. Note that the symmetry
$(\lambda,k) \to (-\lambda,k+\pi)$ is clearly observed in the plot.
Importantly, although the external field $F$ in principle provides a means of getting rid of the interaction energy, our results show that this does not happen. 
Instead, the pair remains bound for weak external fields.

\section{Strong field}
As a contrast to the adiabatic regime, we now consider the regime where the external field $F$ is strong, 
such that $F x_0 \gg 1,\, g_c/ x$ is the largest energy scale.
In this regime, the Coulomb force $g_c/x^2(t)$ between the two electrons is only a small correction to the
constant force $F$. 
Furthermore, the amplitude $2/F$ of the free BO is much smaller than the particle separation $x$, such that, $x(t)$ can be considered as approximately constant. 
Thus, the two particles perform practically independent BOs with
frequencies $\omega_1 \approx F - g_c/x_0^2$ and $\omega_2 \approx F +
g_c/x_0^2$, the frequency difference satisfying
\begin{equation}
\frac{\omega_2-\omega_1}{F} \, \frac{x_0}{g_c} \approx \frac{2}{F x_0}\ ,
\label{eqFreqShift} 
\end{equation}
\begin{figure}[bt]
\includegraphics[width=0.8\linewidth]{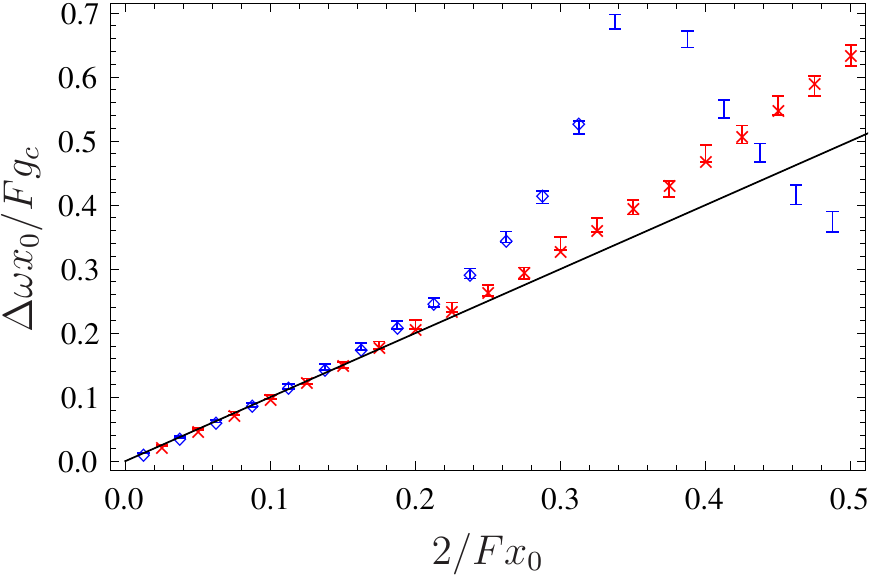}
\caption{(Color online) Frequency difference $\Delta \omega = \omega_2-\omega_1$ obtained from the maxima of the power spectra of
$x_1(t)$ and $x_2(t)$, shown as error bars for $g_c/x_0 = 0.1$ (red, close to the crosses) and
$g_c/x_0 = 1.0$ (blue, close to the diamonds).
The symbols (crosses and diamonds) show the shift from the time-averaged interaction force approximated as $\Delta \omega \approx 2 g_c \langle x^{-2}\rangle$, which deviates
from $\Delta \omega \approx 2g_c x_0^{-2}$ if the conditions $F x_0 \gg 1,\, g_c/ x$ are violated.}
\label{figFreqShift}
\end{figure}%
as plotted in Fig.~\ref{figFreqShift} (solid line). Figure~\ref{figFreqShift}
shows also the frequencies obtained from the full integration of Eqs.\
\eqref{relative_equation_of_motion_dimensionless} for $K_0=k_0=0$, $g_c/x_0 =
0.1$ and $g_c/x_0 = 1.0$, which show good agreement with Eq.~\eqref{eqFreqShift}, as long as
condition $g_c/x^2(t) \ll F$ is well fulfilled. In the opposite case of less separated
particles, the respective oscillations are not independent any more and get
distorted. A good approximation for the frequency shift is then the averaged
interaction force $2 g_c \langle x^{-2}\rangle$ (crosses and diamonds in 
Fig.~\ref{figFreqShift}).

\section{Pairing in higher dimensions}
Finally, we address the phenomenon of pairing due to repulsive interaction in
higher dimensions 
for $F=0$.
Similar to the results shown in Fig.\ \ref{fig1_phasespace}(a) and Fig.\ \ref{fig_2},
we numerically integrate the time evolution under the
three-dimensional generalization of Hamiltonian~\eqref{eqRelHamiltonian} with
initial conditions $\vc{K}_0=0$, $\vc{k}_0=0$, 
and $|\vc{r}_0|$ such that the energy $E_{\rm 0}$ of the relative motion is below, equal to, or above the \emph{effective} half band width 
$\Lambda := 4 \sum_{\mu=1}^{d} |\cos(K_{0,\mu}/2)|$.
Here, we chose $\vc{K}_0 = 0$, such that  $\Lambda = 4d$. 
The orientation of $\vc{r}_0$ is chosen randomly, but the ensuing dynamics is generic. 
\begin{figure}[tb]
\includegraphics[width=\linewidth]{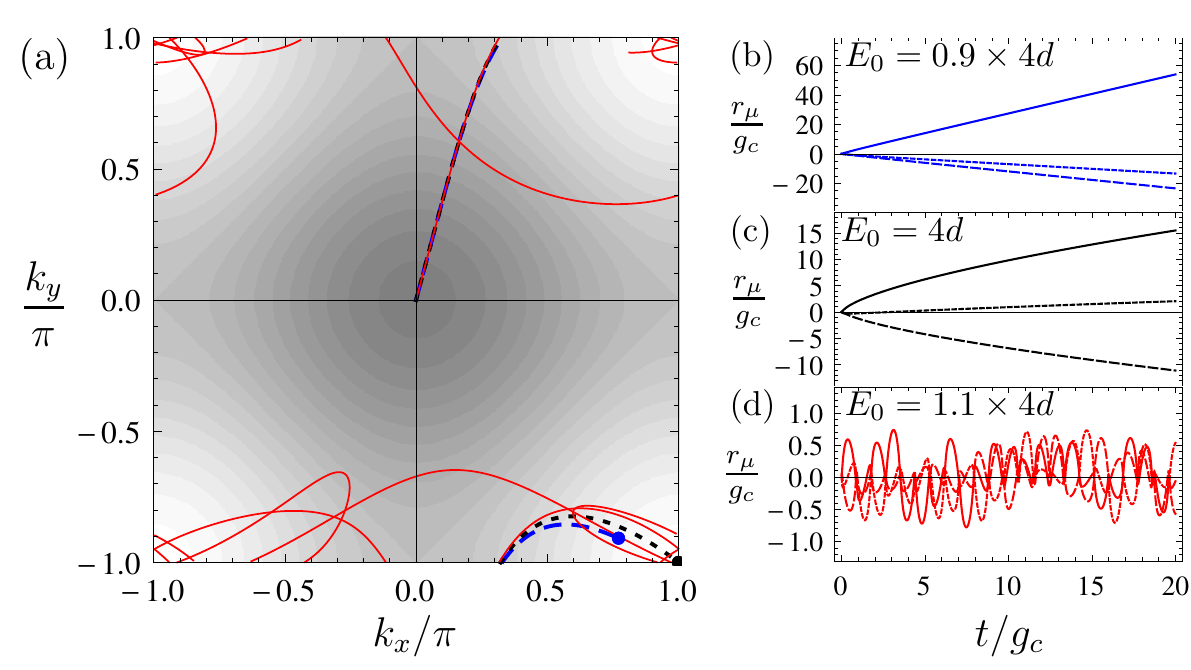}
\caption{(Color online) The pairing transition in $d=3$ dimensions.
The initial conditions are fixed by $\vc{k}_0=0$, $E_{0} = g_c/|\vc{x}_0| - 4d =  \{0.9, 1.0, 1.1\} \times 4d$, and a random but generic orientation of $\vc{r}(0)$. 
(a) projection of the momentum to the $k_x$-$k_y$ plane, the shading indicates the kinetic energy in the first BZ.
In the case $E_0=1.1\times 4d$, $\vc{k}(t)$ is shown until $t/g_c = 4$. In the other cases, the value $\lim_{t\to\infty}\vc{k}$ is marked with a dot.
(b)--(d) real-space time evolution of the Cartesian components $r_\mu$ for the different initial energies.
}
\label{fig3Dpairing}
\end{figure}
The results are shown in Fig.~\ref{fig3Dpairing}.
In the low-energy regime $E_{0} < \Lambda$, $\vc k$ converges to a point that is different from the BZ corner, 
resulting, again, in a finite group velocity and a ballistic separation [see panel (b)]. 
In the limiting case $E_{0} = \Lambda$, $\vc{k}(t)$ converges to one of the corners of the BZ, again with the asymptotics 
$\delta k_j = k_j-n_j \pi  \sim t^{-1/3}$, where the $n_j$ are odd integers.
The real-space asymptotics $x_j(t) \sim t^{2/3}$ is sub-ballistic [see panel (c)].
When the half band width $\Lambda$ is not sufficient to absorb the initial energy $E_0$, the dynamics is quite irregular but the two electrons remain paired due to energetic constraints [panels (a) and (d) of Fig.~\ref{fig3Dpairing}].

If one or several Cartesian components of the initial displacement $\vc{r}_0$
are exactly zero, they remain zero for all times. The effective dimension $d$ is
reduced and the effective half band width $\Lambda=4d$ as well. 
However, these configurations are unstable: the slightest deviation of the initial
orientation grows and the system explores the whole three-dimensional space
at long times.


\section{Conclusions}

A detailed semiclassical analysis of two electrons interacting by the Coulomb
potential in a biased crystal lattice has been presented. The interplay of the
Coulomb force and the external electric field leads to an intricate dynamics
that eventually destroys the harmonic BOs of independent electrons.
Different dynamical regimes of the two electrons were observed, depending on 
their initial separation and the magnitude of the external field. 
When the electrons are far apart, the Coulomb potential may be small as compared to
the external potential and the electrons oscillate with \emph{effective\/} Bloch
frequencies, corresponding to the local electric field. 

If the external field is weak, 
then the electrons either separate without oscillations or, when they are sufficiently close to each other in the beginning, 
they oscillate in the crystal lattice due to the Coulomb field and form a bound
state. 
The reason for this pairing to occur is the finite band width in the tight-binding lattice.
In order to separate the particles, the initial energy $E_0$ has be converted to kinetic energy, which, however, is bounded by the half band width $\Lambda$. 
Thus, the separation of the particles is energetically forbidden if $E_0 > \Lambda$. Then, in dimensions greater than one, $\vc{k}(t)$ performs a kind of unbounded random walk, resulting in aperiodic dynamics of the particle separation $\vc{x}(t)$ in real space. 
We have shown that 
in this situation the role of the external potential is negligible and their dynamics is governed by the Coulomb interaction. 


\acknowledgments

Work in Madrid was supported by MICINN (project MAT2010-17180). Research of
C.G.\ was supported by a PICATA postdoctoral fellowship from the Moncloa Campus
of International Excellence (UCM-UPM). R. P. A. L.  would like to thank FAPEAL
(Alagoas State Research Agency)  via project PRONEX, CNPq, and FINEP (Brazilian
Research Agencies) for partial financial support.


\bibliography{references}

\end{document}